\documentclass[sigconf,nonacm]{acmart}

\AtBeginDocument{%
  \providecommand\BibTeX{{%
    \normalfont B\kern-0.5em{\scshape i\kern-0.25em b}\kern-0.8em\TeX}}}

\setcopyright{none}
\copyrightyear{}
\acmYear{}
\acmDOI{}

\acmConference[]{}
\acmBooktitle{}
\acmPrice{}
\acmISBN{}

\usepackage{subcaption}
\begin{document}

\title[``It was hard to find the words'': An Autoethnographic Diary Study of Smart Home Cyber Security]{``It was hard to find the words'': Using an Autoethnographic Diary Study to Understand the Difficulties of Smart Home Cyber Security Practices}

\author{Sarah Turner}
\email{slt41@kent.ac.uk}
\orcid{0000-0003-1246-1528}
\author{Jason R.C. Nurse}
\email{j.r.c.nurse@kent.ac.uk}
\orcid{0000-0003-4118-1680}
\author{Shujun Li}
\email{s.j.li@kent.ac.uk}
\orcid{0000-0001-5628-7328}
\affiliation{%
  \institution{Institute of Cyber Security for Society (iCSS) \& School of Computing\\University of Kent}
  \city{Canterbury}
  \state{Kent}
  \country{UK}
}


\begin{abstract}
This study considers how well an autoethnographic diary study helps as a method to explore why families might struggle in the application of strong and cohesive cyber security measures within the smart home. Combining two human-computer interaction (HCI) research methods --- the relatively unstructured process of autoethnography and the more structured diary study --- allowed the first author to reflect on the differences between researchers or experts, and everyday users. Having a physical set of structured diary prompts allowed for a period of ``thinking as writing'', enabling reflection upon how having expert knowledge may or may not translate into useful knowledge when dealing with everyday life. This is particularly beneficial in the context of home cyber security use, where first-person narratives have not made up part of the research corpus to date, despite a consistent recognition that users struggle to apply strong cyber security methods in personal contexts. The framing of the autoethnographic diary study contributes a very simple, but extremely powerful, tool for anyone with more knowledge than the average user of any technology, enabling the expert to reflect upon how they themselves have fared when using, understanding and discussing the technology in daily life.
\end{abstract}



\maketitle

\section{Introduction}
\label{sec:introduction}

Internet of Things (IoT) devices within the home setting are increasingly ubiquitous: 2020 in particular saw a huge growth in the purchases of such devices in the UK, attributed in no small part to the amount of time people were required to stay in their homes as part of public health lockdown measures due to COVID-19 \citep{techuk_state_2021}. Home IoT devices\footnote{Throughout this paper, ``home IoT devices" will be used to mean those devices, and those technologies and services that support them, covered in the UK's Code of Conduct for Consumer IoT devices \citep{dcms_code_2018}.} are often left in communal spaces \citep{chalhoub_it_2021}, and, when set up according to the manufacturer's instructions, will collect significant amounts of data about every person that is around them, whether or not those people are aware of it --- or consent to it \citep{koshy_we_2021}. Having an understanding of the risks that these devices pose should be a fundamental part of the purchase and use process, however it is not commonly the case that individuals understand these risks or take steps to manage them \citep{patterson_internet_2021}, or that devices are necessarily designed to make security easy to manage \citep{chalhoub_factoring_2020}.

Cyber security risks that home IoT devices pose are different to those risks that are posed through browsing the Internet on a computer or smart phone \citep{omolara_the_2021}; data collected by these devices can be misused in a number of ways, intentionally or otherwise. In addition, the more devices that are connected to a home network, the greater the threat posed to every part of the home through insecure devices, whether specifically targeted or because of more mundane reasons such as unsupported software \citep{tabassum_i_2019}. Without broad understanding of these risks, users --- and in this case study, we will particularly be considering the family unit --- can be putting themselves in harm's way, unnecessarily. And yet, data and security breaches are commonplace in home IoT devices, both as a result of vulnerabilities in software (see, for example, the list in \citep{srinivas_ten_2020}), but also because users may have failed to use security settings as intended \citep{paul_dozens_2020}.

This case study draws upon two established research methods within the HCI field: that of the diary study, and also, the practice of autoethnographical research. Cyber security is notorious for its poor uptake amongst users; the difficulty of having a coherent and logical cyber security set up within a home increases with the number of users and devices in the household. Families struggle not only to manage device use appropriately, but also to speak and discuss cyber security in meaningful ways, or even use the same language \citep{jones_what_2019}. While researchers study home IoT devices in a professional capacity \citep{williams2019smartwatch}, many of them also use home IoT devices as a user in a personal capacity. Could an autoethnographic diary study, intentionally applying the research lens to the home life of a researcher, help to pick out the specific issues of engagement with the topic? Can it help to create a sense of empathy surrounding the difficulties that non-expert users might have in their daily device use? And what does that mean for how devices are intended to be used, and --- in this case --- kept secure? By analysing ourselves as device users during a period of autoethnographic diary study, can we show where, not only as researchers, but also as device manufacturers or even policy makers, we expect too much --- or too little --- of everyday users? 

This case study details the set up and execution of an autoethnographic diary study,  as a means of exploring the usefulness of first-person, reflexive research into poorly understood areas of digital technology use --- in this case, cyber security habits and practices in the home --- and presents the lessons that have been learned from undertaking it. Although the findings can help academic researchers to consider how to approach not only their topic, but also the users of the particular digital technology, in a more empathetic manner prior to engagement, there are lessons that can also be taken by product designers and also policy makers. All three groups, as experts in their specific topics, can use the structured diary prompts to consider how much of their personal experience is guided by having more knowledge and information than the average user of the product or device. They can also reflect upon the extent to which expert knowledge fails when trying to navigate discussions with family, or friends, or deal with real-life situations. Furthermore, product designers --- in this particular case, in the home IoT space, but also more generally --- may find analysis of what they themselves, their children, friends or family may, or may not, do or know about their product alters their design approach. Policy makers could even apply this method to understand where users may need more education, more support --- or increased regulation or other policy tools to keep them as safe and secure as intended, for example.

This case study starts with a brief review of related work and concepts in Section~\ref{sec:relevant_work}. It goes on to describe the methodology in Section~\ref{sec:method}, and findings in Section~\ref{sec:findings}. Section~\ref{sec:lessons_learned} discusses the lessons that can be taken from the work, prior to conclusions being drawn in Section~\ref{sec:conclusions}.

\section{Relevant Work and Concepts}
\label{sec:relevant_work}

The term ``autoethnographic diary study'' is used here to describe a piece of first-person research exploring the topic in relation to the broader societal and cultural setting, but using the feedback-style recording method of a diary study with multiple participants. It builds upon two research methods: the diary study and autoethnography.

Diary studies are a commonly used method within the HCI and CSCW research communities, as they allow for monitoring of participants' behavior or experiences over an extended period of time, in the moment, rather than relying on recall of events in interview settings. In recent years, diary studies have been used to understand how adolescents (children aged 13-17) and parents manage online harms \citep{wisniewski_dear_2016, mchugh_most_2017, agha_just_2021}, how new parents approach baby wearable technology \citep{wang_quantified_2017}, how children with autism spectrum disorder use mobile applications \citep{putnam_children_2020} and how social groups approach joint privacy and security use of shared applications and devices \citep{watson_we_2020, chalhoub_it_2021}. \citet{garg_when_2019} used a diary study, capturing information from parents with children aged 4-17, on their smart phone and speaker use. Through the entries in the study, they found that there were differences in how families used, managed and limited technology use dependent upon their socio-economic and ethnic status.

The diary studies mentioned above ranged in duration from two weeks to two months, allowing for significant data collection to occur from the participants, both in paper format and using online tools, with reminder capabilities built in.  \citet{watson_we_2020} noted that the ability to track responses online was important, as they needed to chase a number of participants with phone calls to ensure they completed the diary. \citet{hong_using_2020} found that paper allowed for more flexibility in responses --- although participants found managing paper diaries with digital artefacts hard to manage. \citet{putnam_children_2020} found different problems with the diary study method: although the adult participants did not find the method of filling in the diary itself problematic, getting the children involved in the study to participate in a way that generated results to discuss in the diaries proved extremely hard.

The type of personal reflection captured in a journal or diary is core to autoethnographic work, although typically in a much less structured manner than a diary study, capturing any reflections on a specific theme over an extended period of time. \citet{chang_autoethnography_2016} describes autoethnography as autobiographical writing that ``combines cultural analysis and interpretation with narrative details'', and so is particularly relevant when considering the wider use of digital technologies within different areas of society. Such works can be challenging to understand, as they typically raise concerns in relation to the independence, objectivity and generalizability of the method \citep{rapp_autoethnography_2018}. However, the collection of personal thoughts and reflections on a topic for a period of time by a researcher can serve as a lightweight research method that, done well, gives the ability to provide nuanced insights that can outweigh the obvious lack of generalizability~\citep{eschler_critical_2016}. \citet{malinverni_autoethnographic_2016} used autoethnography to determine the importance of how personal values shape their work as researchers, leading to more considered and grounded future research, particularly in the participatory design space.  Analyzing personal use of devices can provide additional levels of empathy towards users and research participants to be taken forward in the design process \citep{okane_gaining_2014, cunningham_autoethnography_2005}; conversely, non-use of devices can also provide insights in that it allows for questioning and re-imagination of use \citep{lucero_living_2018}. Reflecting on the use of closely related duoethnography as a research method, a 2019 paper highlights the importance of using  personal experience to explore the ``interactions between diverse users, devices and data'' in intimate settings \citep{garcia_expanding_2019}; the family unit being one such example.

\section{Methodology}
\label{sec:method}

Following the receipt of ethical approval from our institution's Ethics Committee in August 2020, the first author undertook the research in her own home between 12 August and 31 October 2020. The study mostly focused upon interactions within the first author's immediate family (two children, aged 6 and 3) and husband, although additional interactions with other family members (such as the first author's parents and parents-in-law) that stayed in the home in this period were also captured when relevant. The additional awareness that both the first author and also her husband, being a software engineer, had of cyber security as a topic of household importance was considered to be relevant, as subsequent analysis of the topics raised from the study could determine how many were raised precisely because of this additional awareness. Drawing upon the autoethnographic format, the first author completed the diary entries alone, based upon her interactions and experiences with her family.

\begin{figure*}[ht]
\begin{subfigure}{.33\linewidth}
    \includegraphics[width=\linewidth]{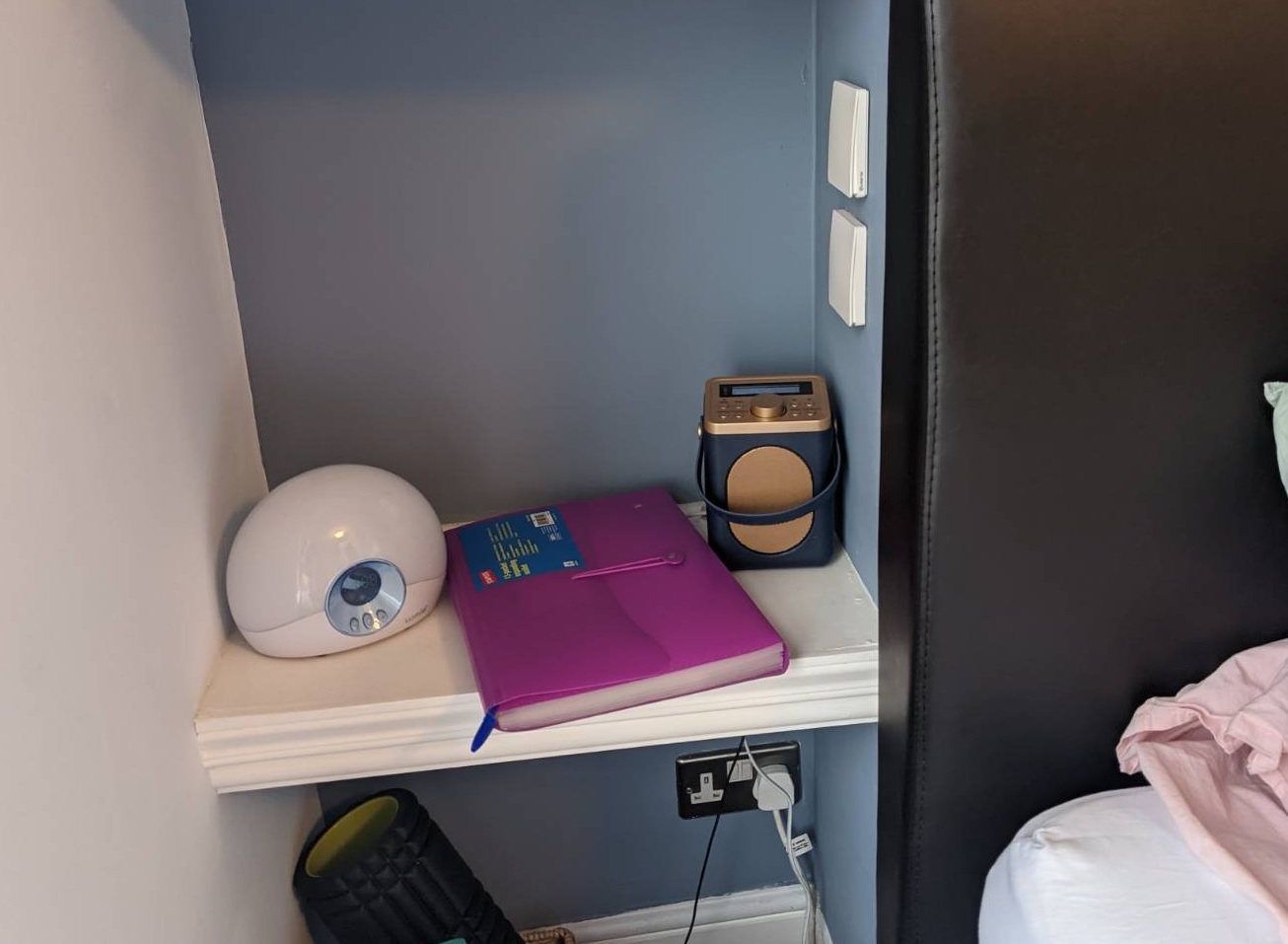}  
    \caption{Bright folder kept by the bed  to prompt \\ daily reflection}
    \label{fig:sub-first}
    \Description[Placement of folder to prompt reflection]{This photograph shows a bright pink folder on a shelf next to the first author's bed, next to a radio and clock: the folder is placed in between the radio and clock, meaning neither can be used without the author seeing the folder.} 
\end{subfigure}
~
\begin{subfigure}{.33\linewidth}
    \includegraphics[width=\linewidth]{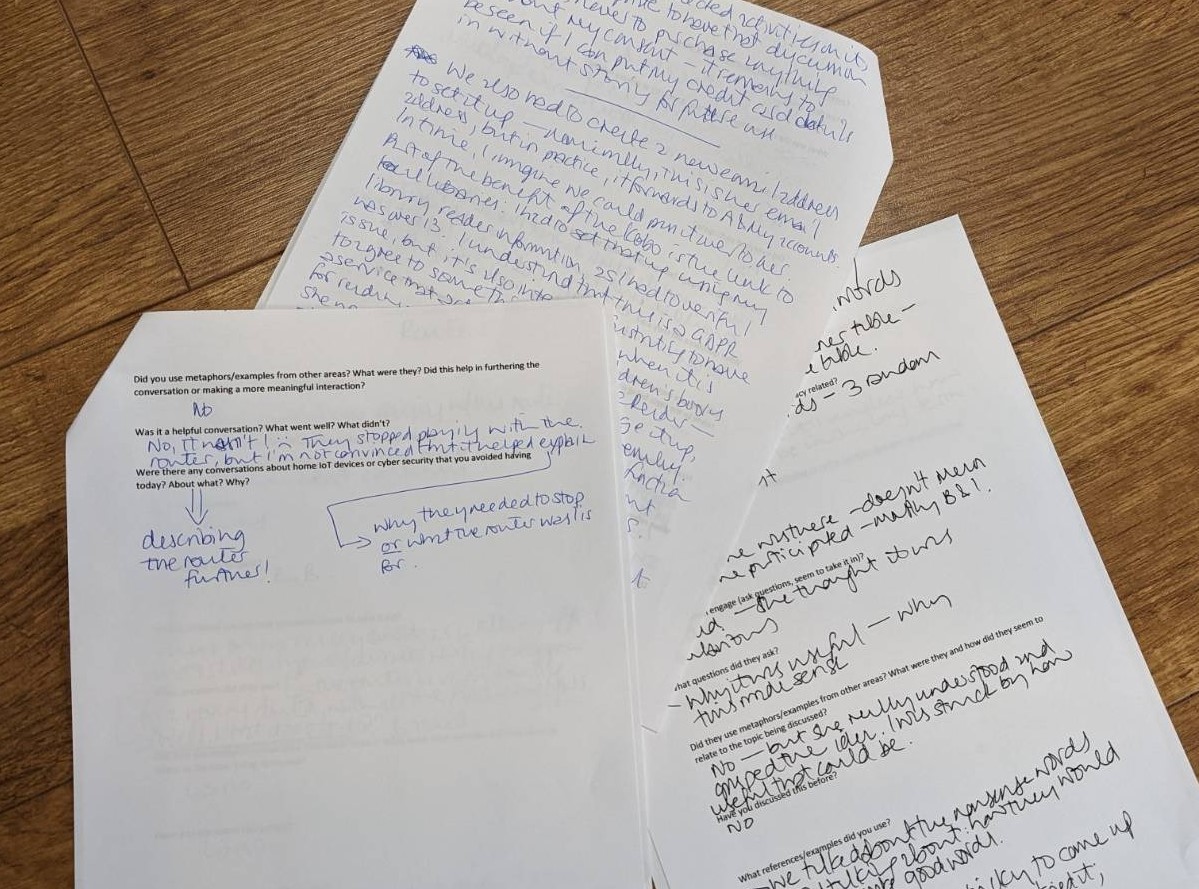}  
    \caption{The pages of the diary prompts, \\incorporating free writing }
    \label{fig:sub-second}
    \Description[Handwritten diary prompts]{This photograph shows three days' diary entries, with handwritten text that covers the backs of two of the sheets of paper, and the third diary entry is one with arrows and other markings around the pre-typed text of the diary prompt sheet.}
\end{subfigure}
~
\begin{subfigure}{.33\linewidth}
    \includegraphics[width=\linewidth]{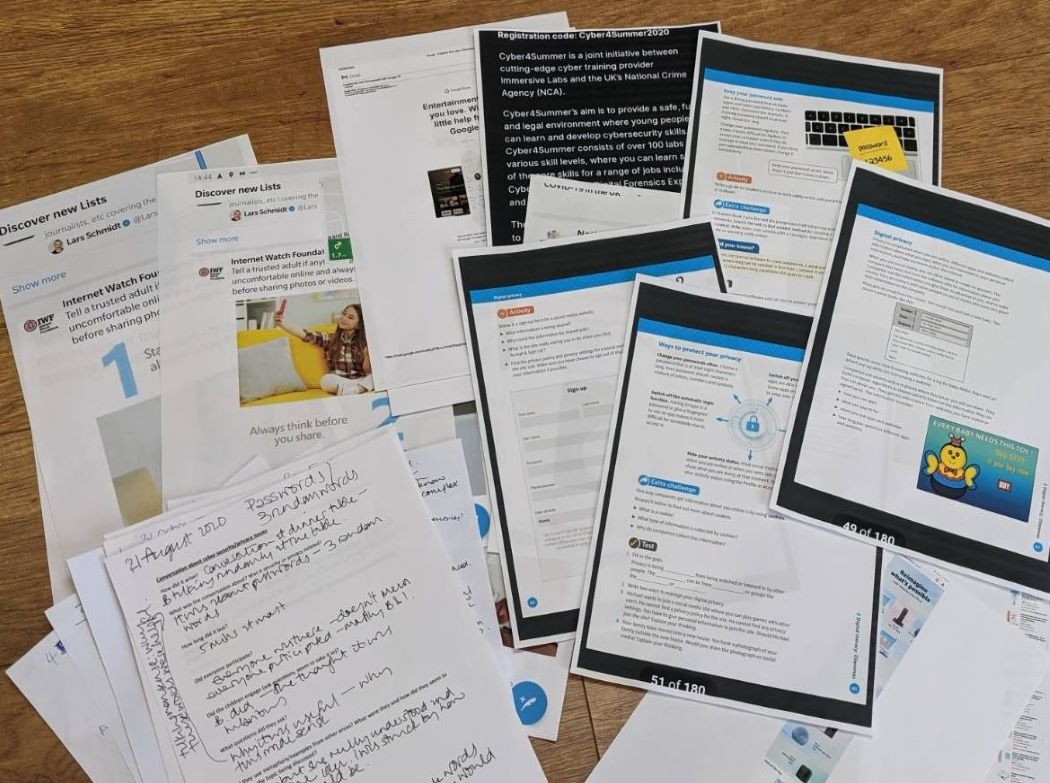}  
    \caption{The full printed diary, with diary entries \\and printed artefacts}
    \label{fig:sub-third}
    \Description[The full diary and artefacts]{A photograph showing the written pages of the diary entries alongside several other printed sheets making up the remaining artefacts of the study.} 
\end{subfigure}
\caption{The diary study process and sample artefacts}
\label{fig:process_artefacts}
\end{figure*}

The diary study topic looked at how the cyber security of home IoT devices was managed and discussed between parents, children and any other relevant individuals within the home. Using a ``feedback'' style of diary \citep{carter_when_2005} in order to elicit broad responses to consider these answers, a set of daily diary prompts posed a series of open-ended questions intended for the first author to reflect on the events of each day pertaining to home IoT device use and cyber security. The questions focused on what was said, done, and what emotions events raised, not dissimilar to the type of responses received in \citep{wisniewski_dear_2016} (for example, ``Was the conversation home IoT device use or cyber security related?'' ``Were there any subjects relating to devices or cyber security of those devices that you avoided talking about today? If so, why?''). The daily diary prompts were printed and kept in a purple folder, along with pens and sufficient additional paper, next to the first author's bed, in order to serve as a visual reminder to log instances at the end of each day (see Figure~\ref{fig:sub-first}). Entries were only to be recorded when there was something of relevance to be captured during the day. For the full list of prompts, see Appendix~\ref{sec:appendix_prompts}. 

The prompts did not change throughout the period, and daily reflections were collected primarily on paper, not electronically (see Figure~\ref{fig:sub-second}). This was for two reasons: the type of activities being considered were unlikely to be done routinely, meaning that using an electronic method for the purposes of eliciting immediate responses through reminders would not be beneficial; also, the use of paper allowed space for more reflection through unstructured feedback \citep{ayobi_flexible_2018}. The hand-written entries were typed up weekly. Any relevant information that was seen online (for example, social media posts) were treated as additional artefacts: they were collated and saved electronically, printed as necessary and analyzed alongside the typed-up diary entries. In the end, the finished diary comprised of written entries, screenshots of social media, school curricula, scans of text books and e-mails, as well as a list of all home IoT devices (and those devices or digital technologies that interacted with the devices) (see Figure~\ref{fig:sub-third}).

Once collated, the complete diary was subjected to thematic analysis \citep{braun_thematic_2006} by the first author. Following \citet{mcdonald_reliability_2019}, it was determined that thematic analysis should only be performed by the first author to preserve the personal and reflexive nature of the research, with broader discussions around the results taking place between all authors.

\section{Findings}
\label{sec:findings}

\subsection{How well did the feedback diary method work?}

\subsubsection{Frequency of reporting}

The pilot diary study lasted 80 days, significantly longer than many documented diary studies, although shorter than many autoethnographic pieces of work. It generated 30 individual diary entries; in addition to the written diary there were 15 screenshots, two e-mails and the list of devices. Many of the diary entries were between 100-300 words in length, with the longest over 700. Despite placing the diary prompts in a convenient location for writing up, the first author felt very aware of the number of days where there was little to nothing to report, based upon the prompts. Electronically saved information often had to be additionally printed to ensure that the thoughts about them were collected at the end of the day.

\subsubsection{What was directly captured in the entries}

The first author found the feedback diary method, in particular with its open-ended questions, helpful as a means of being able to consider and reflect freely upon the situations arising during the diary period. The completed diary covered an extensive range of events, from buying new devices, to discussing reported security breaches and dealing with family device problems, to reflections upon the use of specific software on the first author's smart phone.

Further analysis found that not all of the entries, however, directly contributed to the overall research questions. The entries and artefacts show that two types of cyber security arose in the diary: the ``housework'', of cyber security that is directly applicable within the home, typically relating to things like to device setup, and the ``wider universe'',  reflecting interesting or concerning news stories about cyber security issues that cannot either be directly managed within the home, or that are not directly relevant. In particular, the diary entries allowed the quantification of the time spent considering each type: reading about the cyber security ``wider universe'' appeared in six entries; ``housework'' references to researching new devices prior to purchase, installation of those devices, and device management occurred four times in total.

Those entries that did reflect ``housework'' management of cyber security within the immediate family showed an important element: they were, typically, one-off events. For example, purchasing a new eReader for the first author's eldest child allowed for discussion with the child about setting strong passwords, setting WiFi access, and discussion about how and when books could be purchased or borrowed; this was recorded at the time of setup, and not subsequently.

Once set up in the home, however, questions of device security did not come up. Repeated use of devices seemed to breed familiarity and a level of comfort around its use. When devices were in situ and just functioned as needed, there was no further consideration about the invisible processes in the background that may need further consideration or management. Diary entries discuss long-used devices only in terms of the habitual nature of their use --- both by parents and children, once the device was considered part of the family's setup. ``\textit{The kids are only used to streaming services, and so will often ask to watch programmes via the Chromecast, which allows for useful parental control of what they're watching (as we turn off autoplay). However, this also means that short programs...sees them asking for the next episode almost before the prior one has begun. This is tricky as it can see tired or impatient children grabbing the phone...}'' --- diary entry, 13 September.

The diaries also helped to reflect on the ability of children to consider security, and what that meant for discussions and learning opportunities. There were four detailed discussions about cyber security with the children --- exclusively with the elder child. The younger child was captured in the diary as showing awareness of devices in the home,\footnote{In particular the Google Chromecast that facilitated streaming TV shows.} but had no concept of the need for security. Some of these events allowed for moments of family discussion and collective reflection: for example, password use being mentioned in a television show allowed for a brief discussion of what a password is. In total, passwords were discussed with children three times and unauthorized purchasing once.

\subsubsection{What was indirectly captured in the entries}

The diary prompts did not have questions that required the first author to consider aspects of her role and status, both within the domestic setting and also more broadly in terms of gender and economic status within society. Despite this, both aspects were strongly present within the diary entries, and add a further level of nuance to be included in the analysis. Without a clear understanding of the space in which the researcher inhabits, it may be hard to understand where their experience differs from that of an average user.

Of particular note in this case was the economic status that the first author's family has. The amount spent on new devices and security software in the period led to reflection upon how expensive maintaining appropriate device hygiene can be. The ability to replace those devices that are out of supported software life or use paid-for cyber security software such as password managers may well reflect best practice, but they are options that require sufficient disposable income to make the decision to do so. For many, it could well be a poor decision --- or an impossibility --- to replace otherwise functional devices, or pay for cyber security services in a world where data breaches are common, but obviously tangible downsides are few.

\subsection{How well did the autoethnographic aspect work?}
\label{autoethnographic_aspect}

\subsubsection{An additional level of knowledge}

Using the autoethnographic approach of having only the first author record diary entries was important: in using the reflexive requirement of the study, would it be possible to further deconstruct the reasons why users may typically struggle with managing home IoT device cyber security? The additional level of knowledge held by the first author about requirements and risks associated with device use was clear in a number of entries, giving an idea of privacy and security concerns that might not be considered by those without an interest. Sometimes the entries explored the difficulties of trying to set up devices in ways that are more privacy-preserving and allow for more controlled security: ``\textit{[The eReader] is defaulted to have WiFi on all the time, with limited restrictions on access to the store. Switching off the WiFi results in it warning you that it will cause problems...}'' --- diary entry, 15 October.

Other reported instances of acting out of a heightened interest in security were triggered by external events: for example, trying to find out more about a vulnerability, reported by a technology news site, in microchips used in her and her husband's smart phones:\footnote{\url{https://arstechnica.com/information-technology/2020/08/snapdragon-chip-flaws-put-1-billion-android-phones-at-risk-of-data-theft/}} ``\textit{We really felt that there was little we could do... we'd have to rely on our phone's manufacturer to manage the patching. It unnerved us a bit, in thinking about it, that we found this in specialised press only -–- and certainly not in mainstream news sources. It's tricky having a little bit of knowledge: it often leaves you in a state of uncomfortable inaction...!}'' --- diary entry, 18 August.

\subsubsection{Where knowledge did not help}

Having the written diary entry was helpful to contrast and explore the emotions felt when cyber security was working as expected --- and when it was proving too complex. Negative sentiments were common throughout the diary, with words like ``infuriating'', ``uncomfortable'', ``frustrated'' and ``frustrating'' occurring three times each (in some 6,600 words of the full diary). Dealing with situations that were unresolvable, or that required significant time, knowledge and investment was hard --- even when, as in the first author's case, there was an interest in having the most appropriate security setup at home.

More positive words were less common --- ``amazing'' occurred once, ``benefit'' and ``excellent'' twice each. Interestingly, these more positive records related to the potential use of devices, not aspects associated with security --- there were, in fact, no records commenting that a device's security ostensibly worked. These entries again, help to underline the types of experiences that stick in the mind when using devices as part of life: the first author was inclined to think about security and go out of her way to apply techniques and settings that she knew of, and even then, security activities were framed negatively within the diary. 

The diary entries recorded a number of instances where having cyber security knowledge did not actually help resolve the situation at hand. This was particularly the case when trying to help or communicate about cyber security issues with others.  Trying to help a relative manage some unusual activity on a computer should have been an opportunity to help them walk through and improve their cyber security knowledge and use.  Instead, the relative was so overwhelmed by the situation, and happy once their bank had confirmed no financial loss had occurred as a result of the activity, that they did not listen further. ``\textit{What struck us [first author and her husband] was the complete lack of understanding, backed up with a defensiveness about cyber security practices...Almost everything we tried – both in terms of explanations of mitigating steps, and practically looking at and reviewing the devices – failed.}'' --- diary entry, 17 August.

Similarly, the diary entries showed how having knowledge about how to make devices safe did not help when trying to explain why it was necessary to the children, even when they were keen to listen. The concepts were too hard and too abstract. For example, the children were particularly interested in the first author's new smart phone: ``\textit{It was hard to find the words to explain why I had replaced it – I wanted them to understand that phones are only expected to have a life of around 3 years, but at the same time....they won't understand it! Not happy with the words that fell out of my mouth (`because...it could be dangerous.'). Not that they prodded further – they just loved that the cover wasn't black... They now don't care.}'' --- diary entry, 15 October. The smart phone had been replaced as it had reached the end of its supported life: without guaranteed software updates, it could pose a security risk. The complexity of these ideas, coupled with the uncertainty of the risk (it could pose a risk, should there be a particular set of circumstances), made it a conversation too difficult to have.

Similarly, when the elder child asked what the router did, the best the first author and her husband could do was say ``\textit{well, it's how the Internet comes into the house}'', which ``\textit{felt useless even as we said it...}''. Even if the words were there, the attention span of children for such discussions is extremely limited --- the first author concluded this entry in the diary with a feeling of relief at how quickly the child ``\textit{showed little interest...}'' --- diary entry, 12 September.  One of the artefacts collected alongside the diary entries was the elder child's school curriculum for the year, which detailed the computing skills to be taught during the year. A combination of using a computer (``\textit{how to use or navigate with a mouse}'') and learning about ``\textit{the dangers that the Internet can portray}'' made it clear how ubiquitous computing and the concepts associated with it are not something that the children can hope to learn about at school alongside encountering it at home.
 
\section{Lessons Learned}
\label{sec:lessons_learned}

The autoethnographic diary study method has not previously been used as a way of considering how users of home IoT devices manage cyber security in their homes, and with their families, despite the potential for reflection on day-to-day issues. Although the topic of this study was cyber security, it could be applied to any situation where adoption of understanding of a digital technology is poorer than designers or researchers would hope.

Undertaking a diary study with prompts allowing free-form responses, and the addition of any relevant artefacts, enabled the first author to reflect upon why users find cyber security difficult and unimportant and consider specific reasons for the lack of engagement. The use of a reflexive diary by someone with an expert understanding of what can and should be done to use a digital technology of any kind as intended can be important to show where the process might fail, or to understand those users who are less interested or aware of the steps that might be necessary. Such a method can be powerful in helping not only researchers, but also device designers and policy makers, make recommendations, and base their actions in the mundane of the everyday situation. In particular:

\begin{itemize}
    \item The frequency with with topics arise, and the emotions they generate can help to understand how often a non-expert user might consider the issues, and whether they may actively avoid processes or activities that feel uncomfortable or unpleasant.
    \item Analysis of how and why a topic arises in the diary entry provides some ability to consider how being an expert affects being a user of a device. In particular, allowing for free writing and the addition of artefacts helps to show how and where topics arise: are they from situations and venues non-expert users would encounter?
    \item Analyzing being a user can point out where being an expert does not help. Things remain hard, unexpected, or impossible, even for experts: learning from these experiences is helpful to understand the limits of what users should be expected to endure.
    
\end{itemize}

Below, we go into more detail on each of these points.

\subsection{The frequency of reports}

The inability of the first author to make daily diary entries felt like a concern, when analysis was performed. As recorded, in the 80 days of the study, the first author produced 30 diary entries; a small number compared with similar studies with more participants such as \citet{garg_when_2019}, where the average participant entry rate over an eight-week period was 110, when asked to record information about all types of device use. Furthermore, only six of these entries captured active discussions with the immediate family about cyber security in the context of device use. However, in this respect, the autoethnographic diary method provides the person undertaking the study with a helpful guide as to when and how the topic fits into everyday life. If the individual performing the diary study, who has more interest and specific knowledge than an average user, reports infrequently, this in itself helps to get into the mindset of an average user and stops assuming a level of engagement that may not exist. In this case, for example: if cyber security is only managed at key points of device use, how do you ensure that those brief windows of time are maximized for the best security setup?

\subsection{How and why diary entries arise}

Deciding to use diary prompts requiring answers written on paper, rather than through an online system, facilitated what has been referred to as ``writing as thinking'' \citep{oatley_writing_2008}. When there is something to report, having no limitations on the responses allows for a more reflexive experience, a process referred to as ``critical subjectivity'' in \citep{garcia_expanding_2019}, even if some of the entries end up being outside of the topic of interest at the point of analysis. In particular, when considering a concept that is not widely understood by an average user --- such as cyber security --- the process of writing about instances of dealing with the concept as a more informed researcher helps to understand whether it is reasonable for an average user to consider it too. The majority of diary entries in this case covered wrestling with considerations that came as a result of having researched, and being concerned about, the area for a number of years, and did not spend as much time upon non-specific actions to be taken in the home. If the average user is unlikely to take action, to, for example, limit device use to stop additional data collection, what policy measures might be needed to keep such data safe and used appropriately?

Being able to add in artefacts was another benefit of having a relatively unstructured reporting setup. As previously reported by \citet{hong_using_2020}, keeping artefacts exclusively digitally was not practical for ensuring inclusion and consideration in the wider diary entries, so needed to be printed to ensure this happened. The artefacts were of particular value, however, in bringing the outside world into the home, and reminding the first author of the wider cyber security environment. Again, the chance nature of seeing news items, or social media posts should remind the individual performing the diary study of their particular framing of the world --- would an average user see these posts, or regularly read the news sources that the person performing the diary study considers part of their everyday life?

\subsection{Learning from unexpected and hard things}

Even though performed in a period of enhanced social distancing measures as a result of the COVID-19 pandemic, the diary study allowed for reporting not only of interactions with the first author's nuclear family, but also gave an interesting insight into how the external world encroaches into the home. Giving space to explore the unexpected events in a home setting can prepare researchers, designers and policy makers to think more broadly about the context that they are working in. Prior research has already shown that individuals can find negotiating shared security difficult, even when there is prior agreement as to the importance \citep{watson_we_2020}, and that bystanders pose a particular set of security questions when considering home IoT devices. However, the lived difficulty of these situations may not be truly understood. To consider two examples from the diary study here: first, the situation where the relative may or may not have had a compromised device with access to the first author's home network allowed for exploring the difficulties of acting positively in emotive situations. If the victim of a security issue acts defensively, there is little that anyone else can do, even with a perfect knowledge of the theoretical steps to take.

Second, involving young children is hard. We know that adults and children often use different languages to talk about cyber security \citep{jones_what_2019}, but when the concepts are too complex or too abstract for either the adult to explain or the child to understand, how is that knowledge transfer expected to happen? The diary entries provided space for reflection on how frail the concept of security within the home could be, and that hoping for users to manage this themselves is hard. The artefacts --- school curriculum and text books --- helped to put the inability to talk about this with the children into context: as much as the first author and her husband could not find the words for the cyber security issues they tried to discuss with their children, so the educational system does not set children up to learn about it in enough depth. It also underlines that even when children have parents who understand the reasons for and means of promoting good cyber security in the home, they will not necessarily find the opportunity for discussion and learning at home. This is particularly important to consider from a policy perspective: home IoT devices are increasingly pervasive, yet it is not reasonable to consider that parents have the correct knowledge or vocabulary to discuss safe and secure use of such devices with their children, yet the risks of such device use is not being taught in schools in the UK in any substantive way \citep{department_for_education_national_2013}.

\section{Conclusions}
\label{sec:conclusions}

This case study reports upon the use of an autoethnographic diary study examining the ways the first author's family manage and discuss the cyber security of home IoT devices. Autoethnographic studies and diary studies with multiple participants are relatively common research methods within the HCI field: this case study combined the format of feedback-style diary entries with the reflexive nature of autoethnography. Although autoethnographic work is not generalizable, it was hoped that the reflexivity afforded by such a study might help to further not only understanding why cyber security is poorly understood and managed, but whether the process of performing the diary study could be helpful in understanding where the role of a researcher (or product designer or policy maker, for example), and the role of a user, differs.

The first author found that the diary method, created as it was, to be hand written at the end of every day, allowed for significant opportunities for ``thinking as writing'', unpackaging not only the role of the researcher and the role of the user, but also the complexity of emotions and language around the topic. Having the physical diary entries allowed for analysis of not only the words and language used, and the situations that such language was used, but also for the types of entries, and the frequency of events that were recorded. 

This was particularly valuable in approaching the topic of cyber security, where, despite a consistent recognition that users struggle to apply strong cyber security methods, first-person narratives can help explain the difficulties that even competent users can have with applying good practices in the real world. These findings help to show, in this instance, where cyber security is, and is not, important in a family setting, which can help to frame considerations for not only future research, but also for manufacturers and policy makers. As such, the autoethnographic diary study, despite its simple premise, could be an effective means of providing an expert individual with the reflexive analysis required to unpick problems where users do not act as hoped, by allowing the space for reflection of what it is reasonable for users to know and do, based on the individual's own experiences and reactions.

\appendix

\section{Daily Diary Prompts}
\label{sec:appendix_prompts}

\begin{table}[ht]
	\caption{Daily Diary Prompts}
	\label{tab:diary_prompts}
	\begin{tabular}{p{\linewidth}}
		\toprule
		\textbf{Prompt questions: home IoT device use and cyber security discussions} \\ \midrule
		How did it arise? \\
		Was the conversation home IoT device use or cyber security related? \\
		How long did it last? \\
		Did everyone participate? \\
		Did the children engage (ask questions, seem to take it in)? \\
		What questions did they ask? \\
		Did they use metaphors/examples? What were they and how did they seem to relate to the topic being discussed? \\
		Have you discussed this before? \\
		Did you ask any questions? \\
		Did you use metaphors/examples from other areas? What were they? \\
		Did this help in furthering the conversation or making a more meaningful interaction? \\
		Did you refer to anything else? \\
		Was it a helpful conversation? What went well? What didn’t? \\
		Were there any conversations on digital technologies or cyber security that you avoided having today?  - About what? Why?\\
		\bottomrule
	\end{tabular}
\end{table}

\bibliographystyle{ACM-Reference-Format}
\bibliography{main}

\end{document}